\documentclass[12pt]{article}

\ifx\pdfoutput\undefined
\usepackage[dvips,bookmarks]{hyperref}
\else
\usepackage{hyperref}
\fi
\hypersetup{colorlinks=false,bookmarksopen,bookmarksnumbered,citecolor=blue,
   pdfstartview=FitH}

\usepackage[dvips]{graphicx}
\usepackage{latexsym}
\usepackage{amssymb,amsfonts,amsmath}
\usepackage{graphicx} 
\usepackage{indentfirst}

 \usepackage{bbm}

\parskip=\medskipamount

\newcommand {\cD}{{\cal D}}
\newcommand {\cE}{{\cal E}}

\newcommand {\cL}{{\cal L}}

\newcommand {\cN}{{\cal N}}

\def\a{\alpha}
\def \bi{\bibitem}

\def\b{\beta}

\def\d{\delta}

\def\f{\phi}
\def\g{\gamma}

\def\j{\psi}

\def\l{\lambda}
\def\m{\mu}

\def\q{\theta}

\def\s{\sigma}

\def\u{\upsilon}

\def\z{\zeta}

\def\J{\Psi}
\def\L{\Lambda}

\def\S{\Sigma}
\def\U{\Upsilon}

\def\rd{{\rm d}}

\newcommand{\ad}{{\dot{\alpha}}}                           
\newcommand{\cDB}{{\bar\cD}}                            

\newcommand{\hf}{\frac12}

\newcommand{\be}{\begin{equation}}
\newcommand{\ee}{\end{equation}}
\newcommand{\bea}{\begin{eqnarray}}
\newcommand{\eea}{\end{eqnarray}}
\newcommand{\non}{\nonumber}
\newcommand{\1}{{\underline{1}}}
\newcommand{\2}{{\underline{2}}}

\newcommand{\bm}[1]{\mbox{\boldmath$#1$}}

\def\double #1{#1{\hbox{\kern-2pt $#1$}}}

\newcommand{\gd}{{\dot\g}}

\begin{document}

\begin{titlepage}

\begin{flushright}
July, 2008\\
\end{flushright}
\vspace{5mm}

\begin{center}
{\Large \bf  On  $\bm{\cN=2}$ supergravity and projective superspace: \\
 Dual formulations
}
\end{center}

\begin{center}

{\large  
Sergei M. Kuzenko\footnote{{kuzenko@cyllene.uwa.edu.au}}
} \\
\vspace{5mm}

\footnotesize{
{\it School of Physics M013, The University of Western Australia\\
35 Stirling Highway, Crawley W.A. 6009, Australia}}  
~\\

\vspace{2mm}

\end{center}
\vspace{5mm}

\begin{flushright}
{\it Dedicated to Professor I. L. Buchbinder}\\
{\it On the Occasion of His 60th Birthday}
\end{flushright}

\begin{abstract}
\baselineskip=14pt
The superspace  formulation for four-dimensional $\cN=2$ matter-coupled supergravity 
recently developed in \cite{KLRT-M} makes use of a new type of conformal compensator
with infinitely many off-shell degrees of freedom:
the so-called covariant weight-one polar hypermultiplet.  In the present  note we prove 
the duality of this formulation to the known minimal $(40+40)$ off-shell realization 
for $\cN=2$ Poincar\'e supergravity involving the improved tensor compensator. Within the latter 
formulation, we present new off-shell matter couplings realized in terms of  covariant weight-zero
polar hypermultiplets. We also elaborate upon the projective superspace description 
of vector multiplets in $\cN=2$ conformal supergravity. An alternative superspace representation 
for locally supersymmetric chiral actions is given. We present a model 
for massive improved tensor multiplet with both ``electric''  and ``magnetic'' types
of mass terms. 
\end{abstract}
\vspace{1cm}

\vfill
\end{titlepage}

\newpage
\renewcommand{\thefootnote}{\arabic{footnote}}
\setcounter{footnote}{0}



\section{Introduction}
\setcounter{equation}{0}

Recently, we have developed the superspace formulation 
for four-dimensional $\cN=2$ matter-coupled supergravity \cite{KLRT-M}, 
extending the earlier construction for 5D $\cN=1$  supergravity \cite{KT-Msugra1,KT-Msugra3}.
From the purely geometrical point of view, this approach makes use of Grimm's curved 
superspace geometry \cite{Grimm},
which is perfectly suitable to describe $\cN=2$ conformal supergravity 
and has a simple relation to Howe's superspace formulation \cite{Howe}.
Kinematically, matter fields in \cite{KLRT-M} are described in terms of covariant projective 
supermultiplets which are curved-space versions of  the superconformal 
projective multiplets \cite{K-hyper2} living in rigid projective superspace
\cite{KLR,LR}. In addition to the local $\cN=2$ superspace
coordinates  $z^{{M}}=(x^{m},\q^{\mu}_i,{\bar \q}_{\dot{\mu}}^i)$,
where $m=0,1,\cdots,3$, $\mu=1,2$, $\dot{\mu}=1,2$ and  $i=\1,\2$, 
such a supermultiplet, $Q^{(n)}(z,u^+)$, depends on auxiliary  
isotwistor  variables $u^{+}_i \in  {\mathbb C}^2 \setminus  \{0\}$, 
with respect to which $Q^{(n)}$ is holomorphic and homogeneous,
$Q^{(n)}(c \,u^+) =c^n \,Q^{(n)}(u^+)$, on an open domain of ${\mathbb C}^2 \setminus  \{0\}$
(the integer parameter $n$ is called the weight of   $Q^{(n)}$).
In other words, such superfields are intrinsically defined in  ${\mathbb C}P^1$.
The covariant projective  supermultiplets are required to be 
annihilated by half of the supercharges\footnote{In the rigid 
supersymmetric case, such constraints  in isotwistor superspace
${\mathbb R}^{4|8}\times {\mathbb C}P^1$  
were introduced  first by Rosly 
\cite{Rosly},   and later  by the harmonic \cite{GIKOS,GIOS} 
and projective \cite{KLR,LR} superspace practitioners.}  
\be
\cD^+_{\a} Q^{(n)}  = {\bar \cD}^+_{\ad} Q^{(n)}  =0~, \qquad \quad 
\cD^+_{ \a}:=u^+_i\,\cD^i_{ \a} ~, \qquad
{\bar \cD}^+_{\dot  \a}:=u^+_i\,{\bar \cD}^i_{\dot \a} ~,
\label{ana}
\ee  
with $\cD_{{A}} =(\cD_{{a}}, \cD_{{\a}}^i,\cDB^\ad_i)$ the covariant superspace 
derivatives.

In the approach of \cite{KLRT-M}, the dynamics of supergravity-matter 
systems is described by a  locally supersymmetric action of the form:
\bea
S&=&
\frac{1}{2\pi} \oint (u^+ \rd u^{+})
\int \rd^4 x \,{\rm d}^4\q {\rm d}^4{\bar \q}
\,E\, \frac{{\bm W}{\bar {\bm W}}\cL^{++}}{({\bm \S}^{++})^2}~, 
\qquad E^{-1}= {\rm Ber}(E_A{}^M)~,
\label{InvarAc}
\eea
where
 \be
{\bm \S}^{++}:=\frac{1}{4}\Big( (\cD^+)^2 +4S^{++}\Big){\bm W}
=\frac{1}{4}\Big( ({\bar \cD}^+)^2 +4\widetilde{S}^{++}\Big){\bm {\bar W}}
=\S^{ij}u^+_i u^+_j~.
\label{Sigma}
\ee
Here the Lagrangian $\cL^{++}(z,u^+)$ is a covariant real projective 
multiplet of weight two, $\bm W(z)$ is the covariantly chiral field strength 
of an Abelian vector multiplet (i.e. the first superconformal compensator),
such that the body of  $\bm W(z) $ is everywhere non-vanishing,
$S^{++}(z,u^+)=S^{ij}(z)u^+_i u^+_j$ and 
$\widetilde{S}^{++}(z,u^+)={\bar S}^{ij}(z)u^+_i u^+_j$ 
are special dimension-1 components of the supertorsion; see \cite{KLRT-M}
for more detail. The action (\ref{InvarAc}) can be shown to be invariant 
under the supergravity gauge transformations, and is also manifestly 
super-Weyl invariant.
It can also be rewritten in the equivalent form
\bea
S&=&
\frac{1}{2\pi} \oint (u^+ \rd u^{+})
\int \rd^4 x \,{\rm d}^4\q {\rm d}^4{\bar \q}
\,E\, \frac{\cL^{++}}{ S^{++} \widetilde{S}^{++}}~ 
\label{InvarAc2}
\eea
in which, however, the super-Weyl invariance is not manifest.

In \cite{KLRT-M}, we presented a family of supergravity-matter systems
in which the matter hypermultiplets are described by 
covariant weight-zero polar multiplets\footnote{We follow the terminology 
introduced in \cite{G-RLRvUW} in  the rigid supersymmetric case.},
and the second superconformal compensator is identified with 
a covariant weight-one polar multiplet.
This is a new type of supergravity compensator, although 
it is related to  the $q^+$-hypermultiplet compensator which emerges 
within the harmonic superspace approach \cite{GIOS2,GIOS} to 4D $\cN=2$
supergravity, e.g., in the sense of \cite{K-double}.
In the present paper, we wish to study duality of the supergravity 
formulation given in \cite{KLRT-M} to (one of) the minimal $(40+40)$ off-shell formulations 
for $\cN=2$ Poincar\'e supergravity  constructed in the 1980s.
These formulations are obtained by coupling the minimal field representation 
with $32+32$ off-shell degrees of freedom
 \cite{BS1} (that is the Weyl multiplet  \cite{deWvHVP,BdeRdeW,deWLVP}  
coupled to an Abelian vector multiplet, the latter being the first superconformal compensator) 
to various off-shell versions for the second compensator with $(8+8)$ degrees of freedom.
These include:
(i) the ``standard'' minimal realization with a nonlinear  multiplet  \cite{FV,deWvHVP};
(ii) the alternative  formulation involving  an off-shell hypermultiplet 
with intrinsic central charge \cite{alternative};
(iii) the new minimal realization with an improved tensor multiplet \cite{deWPV}.
Superspace realizations for these supergravity formulations have been studied, 
e.g., in \cite{BS,Howe,GIOS}.
Our analysis will be specifically concerned with the duality between 
the third minimal Poincar\'e supergravity \cite{deWPV} and the supergravity 
formulation given in \cite{KLRT-M}. The point is that the former 
is known to be analogous to the new minimal $\cN=1$ Poincar\'e supergravity  \cite{new}.
We are going to demonstrate below that the projective-superspace formulation 
\cite{KLRT-M} is analogous to the old minimal $\cN=1$ Poincar\'e supergravity 
\cite{WZ-s,old}.\footnote{The ``standard'' minimal formulation
for $\cN=2$ Poincar\'e supergravity  \cite{FV,deWvHVP}
is analogous to the $\cN=1$ non-miminal supergravity \cite{non-minimal,SG}, 
and thus it seems to be hardly useful for practical (e.g., supergraph) calculations.
As to the alternative formulation for $\cN=2$ supergravity \cite{alternative}, 
its best superspace description appears to be achieved within the harmonic 
superspace approach, as worked out in \cite{KT} and reviewed in \cite{GIOS}.}

This paper is organized as follows.
In section 2, we elaborate upon the projective-superspace description 
of Abelian vector multiplets in $\cN=2$ conformal supergravity, 
and present an alternative superspace representation 
for locally supersymmetric chiral actions.
In section 3, we start by describing the supespace realization 
for the improved vector multiplet (both massless and massive)
coupled to conformal supergravity. We then present new off-shell matter couplings
within the third  minimal Poincar\'e supergravity \cite{deWPV}. 
And finally, the duality of such supergravity-matter systems to those
presented in \cite{KLRT-M} is explicitly proved.
In the appendix we provide four equivalent forms for 
the free $\cN=2$ vector multiplet action in conformal supergravity. 
Our two-component notation and conventions follow \cite{BK}, 
and these are almost identical to those adopted in \cite{WB}.

\section{Vector multiplets in conformal supergravity}
\setcounter{equation}{0}

In this section, we   elaborate upon the projective-superspace description 
of Abelian vector multiplets in conformal supergravity, building on
\cite{KLRT-M,KT-M-AdS}, and also propose an alternative superspace realization 
for $\cN=2$ locally supersymmetric chiral actions. 
Following the supergravity conventions adopted  \cite{KLRT-M}, an Abelian vector multiplet
is described by  its field strength $W(z)$ which is covariantly chiral, 
\be
{\bar \cD}^{\ad }_i W =0~,
\ee
and obeys  the Bianchi identity
\bea
\S^{ij}:=\frac{1}{4}\Big(\cD^{\g(i}\cD_\g^{j)}+4S^{ij}\Big)W
&=&
\frac{1}{4}\Big(\cDB_\gd^{(i}\cDB^{ j) \gd}+ 4\bar{S}^{ij}\Big)\bar{W}=:{\bar \S}^{ij}
~.
\label{vectromul}
\eea
Under the infinitesimal super-Weyl transformation,   $W$ varies as 
\be
\d_{\s} W = \s W~,
\label{Wsuper-Weyl}
\ee
with $\s$ an arbitrary covariantly chiral scalar.
The super-Weyl transformation of $\S^{ij}$
is 
\be
\d_{\s} \S^{ij} = \big(\s +\bar \s \big)  \S^{ij} ~.
\label{S(++)super-Weyl}
\ee

The vector multiplet can also be described in terms of a gauge prepotential\footnote{In 
the harmonic superspace approach \cite{GIKOS,GIOS}, one uses a different gauge prepotential
which is globally defined on $S^2={\mathbb C}P^1$.
The explicit relationship between the harmonic and the projective superspace 
formulations is spelled out in \cite{K-double} in the rigid supersymmetric case.} 
which we identify with 
a covariant real weight-zero tropical supermultiplet,  
$V(u^+)$. 
In the north chart of  ${\mathbb C}P^1$ parametrized by the
complex coordinate $\z =u^{+\2}/u^{+\1}$,   the prepotential 
is specified by the following properties:
\bea
\cD^+_{\a} V  = {\bar \cD}^+_{\ad} V  =0~, \qquad
V(u^+)= 
V(\z) =\sum_{k=0}^{+\infty}\z^k \,V_k~,\quad V_k=(-1)^k\bar{V}_{-k}~.~~
\eea
The prepotential is defined modulo  gauge transformations
of the form: 
\be
\d V =\l  + \widetilde{\l}~,
\label{vm-gauge-tr}
\ee
with the gauge parameter $\l(u^+)$ being a covariant weight-zero arctic multiplet, 
and $\widetilde{\l}$ its smile-conjugate (see, e.g., \cite{KLRT-M} for the definition of the 
smile-conjugation), 
 \begin{subequations} 
\bea
\cD^+_{\a} \l  &=& {\bar \cD}^+_{\ad} \l  =0~, \qquad
\l(u^+) = \l(\z) =\sum_{k=0}^{+\infty}\z^k\l_k ~,  
\\
\cD^+_{\a} \widetilde{\l}  &= &{\bar \cD}^+_{\ad} \widetilde{\l}  =0~, \qquad
\widetilde{\l}(u^+) = \widetilde{\l}(\z) =\sum_{k=0}^{+\infty}(-1)^k \z^{-k}\bar{\l}_k 
~.
\eea
 \end{subequations} 

It turns out that the field strength $W$ and its conjugate $\bar W$ are expressed
in terms of the prepotential $V$ as follows \cite{KT-M-AdS}:
 \begin{subequations} 
\bea
W  &=&
-\frac{1}{8\pi}\oint \frac{(u^+\rd u^+)}{(u^+u^-)^2}
\Big( (\cDB^{-})^2
+4\widetilde{S}^{--}\Big)V(u^+)~,
\label{W} \\
\bar{W} &=&-\frac{1}{ 8\pi}\oint
\frac{(u^+\rd u^+)}{(u^+u^-)^2}
\Big( (\cD^{-})^2+4S^{--} \Big)V(u^+)
~,~~~~~~~~~
\label{W-bar}
\eea
\end{subequations} 
where the contour integral is carried out around the origin,
${\bar \cD}^-_\ad = u^-_i{\bar \cD}^i_\a $ and $\cD^-_\a = u^-_i\cD^i_\a$,
$ \widetilde{S}^{\pm \pm}= u^\pm_i u^\pm_j {\bar S}^{ij}$  
and $S^{\pm \pm}= u^\pm_i u^\pm_j S^{ij}$.
Here we have introduced  an additional complex two-vector,  $u^-_i$, which is only subject 
to the condition $(u^+u^-) := u^{+i}u^-_i \neq 0$, and is otherwise completely arbitrary.
The right-hand sides of  (\ref{W}) and (\ref{W-bar})
can be seen to be invariant 
under arbitrary projective transformations of the form:
\be
(u_i{}^-\,,\,u_i{}^+)~\to~(u_i{}^-\,,\, u_i{}^+ )\,R~,~~~~~~R\,=\,
\left(\begin{array}{cc}a~&0\\ b~&c~\end{array}\right)\,\in\,{\rm GL(2,\mathbb{C})}~.
\label{projectiveGaugeVar}
\ee
The representations  (\ref{W}) and (\ref{W-bar}) generalize similar results in 
the 5D $\cN=1$ flat \cite{K-hyper1} and Anti-de Sitter \cite{KT-M-5DAdS}
superspaces.

Using the fact that $V(u^+)$ is a covariant projective supermultiplet of weight zero,
in particular $\cD^+_{\a} V  = {\bar \cD}^+_{\ad} V  =0$, one can show that the right-hand side 
of (\ref{W}) is covariantly chiral \cite{KT-M-AdS}.
The field strength $W$, eq. (\ref{W}), 
turns out 
to be invariant under the gauge transformations
(\ref{vm-gauge-tr}) \cite{KT-M-AdS}.

In accordance with the general results on the super-Weyl transformation
laws of covariant projective multiplets \cite{KLRT-M}, 
the gauge prepotential must be inert under the super-Weyl transformations,
\be
\d_\s V =0~.
\ee
It can be demonstrated that this transformation law
implies the super-Weyl transformation of $W$, 
eq. (\ref{Wsuper-Weyl}).

Let ${\bm V}(u^+)$ be the tropical prepotential for the vector 
multiplet given by the field strengths $\bm W$ and $\bar {\bm W}$
appearing in the supersymmetric action (\ref{InvarAc}).
The dynamics of this vector multiplet can be described by the Lagrangian \cite{KLRT-M}
\bea
\cL^{++}_{\rm vector} = -\hf {\bm V}\,{\bm \S}^{++}~.
\label{vm-lagrangian}
\eea
We wish to express the corresponding action, $S_{\rm vector}$, in terms of 
the prepotential.

Making use of eq. (\ref{W}) gives
\bea
{\bm \S}^{++}(u^+) &=& 
\frac{1}{4}
\Big( (\cD^{+})^2+4{S}^{++}\Big){\bm W}\non \\
&=&-\frac{1}{32\pi} \Big( (\cD^{+})^2+4{S}^{++}\Big)
\oint \frac{(\hat{u}^+\rd \hat{u}^+)}{(\hat{u}^+u^-)^2}
\Big( (\cDB^{-})^2
+4\widetilde{S}^{--}\Big){\bm V}(\hat{u}^+)~.
\eea
Here the expression in the second line does not depend on $u^-_i$,
and the freedom to choose $u^-_i$ can be used to set $u^-_i = u^+_i$. 
This leads to the following representation
\bea
{\bm \S}^{++}(u^+) &=& 
-\frac{1}{32\pi} \Big( (\cD^{+})^2+4{S}^{++}\Big)
\Big( (\cDB^{+})^2+4\widetilde{S}^{++}\Big)
\oint \frac{(\hat{u}^+\rd \hat{u}^+)}{(\hat{u}^+u^+)^2}
{\bm V}(\hat{u}^+) \non \\
&=&-\frac{1}{32\pi} 
\Big( (\cDB^{+})^2+4\widetilde{S}^{++}\Big)
\Big( (\cD^{+})^2+4{S}^{++}\Big)
\oint \frac{(\hat{u}^+\rd \hat{u}^+)}{(\hat{u}^+u^+)^2}
{\bm V}(\hat{u}^+)~,
\eea
which makes manifest the fact that ${\bm \S}^{++}$ is a covariant projective multiplet.
Plugging the expression obtained into the action $S_{\rm vector}$ and then integrating by 
parts gives
\bea
S_{\rm vector}&=& 
\frac{1}{2\pi} \oint (u^+ \rd u^{+})
\int \rd^4 x \,{\rm d}^4\q {\rm d}^4{\bar \q}\,E\, 
\frac{{\bm W}{\bar {\bm W}}\cL^{++}_{\rm vector}}{({\bm \S}^{++})^2}
\non \\
&=&\hf \frac{1}{(2\pi)^2} \oint (u^+_1 \rd u^{+}_1)  \oint (u^+_2 \rd u^{+}_2)
\int \rd^4 x \,{\rm d}^4\q{\rm d}^4{\bar \q}\,E\,
\frac{{\bm V}(u^+_1){\bm V}(u^+_2)}{(u^+_1u^+_2)^2}~.
\label{VM-action}
\eea
This result is a curved-superspace generalization 
of the rigid supersymmetric action for the vector multiplet 
in projective superspace \cite{G-R}. The fact that here we have dealt  
with a particular vector multiplet, is not actually relevant.
In the appendix, we generalize (\ref{VM-action}) to the case of  an arbitrary
vector multiplet.

The description of vector multiplets in terms of their projective prepotentials
may look somewhat exotic. Having this in mind, we would like to demonstrate
its equivalence to the standard formulation, in which the vector multiplet action is given 
as an integral over the chiral subspace, originally presented  in \cite{GSW}
in the case of rigid supersymmetry and then extended to supergravity, e.g.,
in \cite{Muller}. Since the curved-superspace considerations require the use
of a chiral density (see \cite{Muller} and references therein),  which makes
the analysis somewhat lengthy and technical, we restrict ourselves to 
the flat case. Using
the flat-superspace representation 
\bea
{\bm \S}^{++} &=& \frac{1}{4} (D^{+})^2{\bm W}
=\frac{1}{4}({\bar D}^+)^2 {\bar {\bm W}}~,
\eea
$S_{\rm vector}$ can be transformed as follows:
\bea
S_{\rm vector}&=& 
-\frac{1}{4\pi} \oint (u^+ \rd u^{+})
\int \rd^4 x \,{\rm d}^4\q {\rm d}^4{\bar \q}\, 
\frac{{\bm W}{\bar {\bm W}}}{{\bm \S}^{++}}\,{\bm V}\non \\
&=& -\frac{1}{64\pi} \oint (u^+ \rd u^{+})
\int \rd^4 x \,{\rm d}^4\q \,
\frac{ ({\bar D}^-)^2 ( {\bar D}^+)^2 }{(u^+u^-)^2}
\frac{{\bm W}{\bar {\bm W}}}{{\bm \S}^{++}}\,{\bm V} \non \\
&=& -\frac{1}{16\pi} \oint
 \frac{(u^+ \rd u^{+})}{(u^+u^-)^2}
\int \rd^4 x \,{\rm d}^4\q \,
{\bm W}\, ({\bar D}^-)^2{\bm V} 
=\hf \int \rd^4 x \,{\rm d}^4\q \,
{\bm W}^2~.
\eea
In the last line, we have used the flat-superspace version of eq. 
(\ref{W}). 

As a natural generalization, consider a system of $n+1$ Abelian 
vector multiplets described by covariantly chiral field strengths $W_I$, 
where $I=0,1,\dots, n$, and $W_0 = {\bm W}$.
Their dynamics can be described by the Lagrangian \cite{KLRT-M}
\be
\cL^{++} = -\frac{1}{4} {\bm V} \,\Big\{ \Big( (\cD^{+})^2+4{S}^{++}\Big) F(W_I)
+\Big( (\cDB^{+})^2+4\widetilde{S}^{++}\Big){\bar F}({\bar W}_I) \Big\}~,
\label{vector-self}
\ee
with $F(W_I)$ a holomorphic homogeneous function of degree one,
$F(c W_I)=cF(W_I)$.
The construction given  admits an obvious extension to the non-Abelian case.

In the {\it flat-superspace} limit, the  action generated by (\ref{vector-self})
can be represented in the following different but equivalent forms:
\bea
S&=& 
\frac{1}{2\pi} \oint (u^+ \rd u^{+})
\int \rd^4 x \,{\rm d}^4\q {\rm d}^4{\bar \q}\, 
\frac{{\bm W}{\bar {\bm W}}  \cL^{++} }{({\bm \S}^{++})^2 }
=\int \rd^4 x \,{\rm d}^4\q \,
{\bm W}^2 F\Big(\frac{W_I }{ {\bm W}}\Big) +{\rm c.c.}~~~~~
\eea
It should be pointed out that the representation (\ref{vector-self}), which 
describes effective vector multiplet models in supergravity, is a natural 
generalization of that given in \cite{DKT} in the rigid supersymmetric 
case using  harmonic superspace techniques.

The above consideration leads to a new representation for chiral actions
in $\cN=2$ supergravity that avoids any use of the chiral density
(as mentioned earlier, the latter requires some care to 
be explicitly constructed \cite{Muller}). Let $\cL_{\rm c} (z)$ be a covariantly 
chiral scalar superfield, ${\bar \cD}_\ad \cL_{\rm c} =0 $, 
with the super-Weyl transformation 
\be
\d_{\s} \cL_{\rm c} = 2 \s \cL_{\rm c}~.
\ee
For the chiral action $S_{\rm c}$ associated with $\cL_{\rm c}$, we have 
\bea
S_{\rm c}= \int \rd^4 x \,{\rm d}^4\q \, \cE \, \cL_{\rm c} &+& {\rm c.c.}
=\frac{1}{2\pi} \oint (u^+ \rd u^{+})
\int \rd^4 x \,{\rm d}^4\q {\rm d}^4{\bar \q}\, E\,
\frac{{\bm W}{\bar {\bm W}}  \cL^{++}_{\rm c} }{({\bm \S}^{++})^2 }~, \non \\
\cL^{++}_{\rm c} &=&
 -\frac{1}{4} {\bm V} \,\Big\{ \Big( (\cD^{+})^2+4{S}^{++}\Big) \frac{\cL_{\rm c}}{\bm W}
+\Big( (\cDB^{+})^2+4\widetilde{S}^{++}\Big)
\frac{{\bar \cL}_{\rm c} }{\bar {\bm W}} \Big\}~.~~~~~~
\label{chiral-action}
\eea
If $\cL_{\rm c}$ is independent of the vector multiplet described by $\bm V$, 
then one can show, 
using the representations (\ref{W}) and (\ref{W-bar}), that $S_{\rm c}$ 
does not change under an arbitrary variation of the prepotential $\bm V$, 
\be
\frac{\d}{\d \bm V} \cL_{\rm c} =0 \qquad \Longrightarrow \qquad 
 \frac{\d}{\d \bm V} S_{\rm c} =0~.
\ee
The derivation of this result requires transformations similar to those described in the Appendix.

As an example of chiral models,  consider a higher-derivative Lagrangian
of the form: 
\bea
\cL_{\rm c} =\frac{(W^{\a\b}W_{\a\b })^n}{ ({\bm W})^{2n-2}}~,
\eea
with $W_{\a\b}$ the $\cN=2$ super-Weyl tensor.

\section{Dual formulations for matter-coupled supergravity}
\setcounter{equation}{0}

We are prepared for the analysis of  supergravity-matter systems 
and their dualities.\footnote{Examples of duality transformations for rigid 
projective supermultiplets were considered, e.g., in \cite{G-RLRvUW}.}

\subsection{Massless and massive improved tensor multiplets}
The improved $\cN=2$ tensor multiplet\footnote{The  improved $\cN=1$ tensor multiplet
was introduced by de Wit and Ro\v{c}ek \cite{deWR}.
It is a unique superconformal model in  the family of $\cN=1$ tensor multiplet 
models discovered by Siegel \cite{Siegel}.} 
\cite{deWPV,LR2,KLR,GIO}
occurs as a conformal compensator in one of the  off-shell formulations
for 4D $\cN=2$ Poincar\'e supergravity which was developed in \cite{deWPV}
using the $\cN=2$ superconformal tensor calculus.
Here we start by presenting a curved superspace realization for the 
improved tensor multiplet, 
building on the rigid projective superspace formulation for this multiplet
given in \cite{KLR}.

The $\cN=2$ tensor multiplet\footnote{In rigid $\cN=2$ supersymmetry, the off-shell
tensor multiplet was first introduced by Wess \cite{Wess}, and its projective superspace 
realization was given in \cite{KLR}.} 
is described by a covariant real $O(2)$
multiplet $G^{++}$,
\be
G^{++}(u^+)= G^{ij} u^+_i u^+_j~, \qquad 
\overline{G^{ij}}=G_{ij}~, 
\qquad \cD^{(i}_\a G^{jk)} = {\bar \cD}^{(i}_\ad G^{jk)} =0~.
\ee
The Lagrangian for the improved tensor multiplet is
\be
\cL^{++}_{\rm impr.-tensor} =- G^{++} \ln \frac{G^{++}}{{\rm i}\widetilde{ \U}^+{ \U}^+}~,
\label{improved}
\ee
where ${ \U}^{+}(u^+)$ is a covariant weight-one {\it arctic} multiplet, 
and $\widetilde{\U}^+(u^+)$ its smile-conjugated {\it antarctic} superfield.
The action can be seen to be independent of ${ \U}^{+}$ and 
$\widetilde{\U}^+$. Indeed, one can show that 
\bea
\oint (u^+ \rd u^{+})
\int \rd^4 x \,{\rm d}^4\q{\rm d}^4{\bar \q}\,
E\, \frac{{\bm W}{\bar {\bm W}}}{({\bm \S}^{++})^2}\,G^{++} \,\l =0~, 
\eea
for an arbitrary covariant weight-zero {\it arctic} multiplet
$\l(u^+)$. Therefore, the action generated by (\ref{improved}) is 
invariant under gauge transformations ${ \U}^{+} \to {\rm e}^{\l}\,{ \U}^{+}$, 
and  thus ${ \U}^{+}$ can be gauged away.
In other words, ${ \U}^+$ is a purely degree of freedom.
Both $G^{++}$ and ${ \U}^{+}$ are required to possess
non-vanishing expectation values.

The action generated by (\ref{improved}) is manifestly invariant under the super-Weyl
transformations
\be
\d_\s G^{++} = (\s +{\bar \s})G^{++}~, \qquad 
\d_\s {\U}^{+} =\hf (\s+ {\bar \s}) { \U}^+~.
\label{super-Weyl-G}
\ee
These super-Weyl transformation laws of $G^{++}$ and ${ \U}^+$ are 
determined by their off-shell structure \cite{KLRT-M}.
In the flat superspace limit, the Lagrangian (\ref{improved})
provides a {\it manifestly superconformal formulation} for the 
improved tensor multiplet within the 
superconformal formalism given in \cite{K-hyper2}.

One can consider a coupling of $G^{++} $ 
to an Abelian vector multiplet generated by the Lagrangian
\bea
\cL^{++} = -\hf { V}\,{ \S}^{++}
-{ G}^{++} 
\ln \frac{{ G}^{++}}{{\rm i}\widetilde{ \U}^+{\rm e}^{m V}{ \U}^+} ~,
\label{improved-massive-1}
\eea
with $m$ a real parameter.
The corresponding action is invariant under 
the gauge transformations (\ref{vm-gauge-tr}).
This model describes a massive improved tensor multiplet.
In the rigid supersymmetric case, it was introduced 
in \cite{LR2} in terms of $\cN=1$ superfields
(as a model for $\cN=2$ supersymmetric QED), 
then in \cite{Siegel83} in terms of ordinary $\cN=2$ superfields,
and later its description in $\cN=2$ projective superspace
was given \cite{Siegel-curved}. 

Massive two-forms 
naturally appear in four-dimensional  
$\cN=2$ supergravity  theories 
obtained from (or related to) compactifications 
of type II string theory on Calabi-Yau threefolds
in the presence of  both electric and magnetic 
fluxes \cite{LM,DSV}. 
This led to renewed interest 
in $\cN=1$ and $\cN=2$  rigid massive tensor multiplets\footnote{Models for  massive 
$\cN=1$ tensor multiplet  were proposed  for the first time  in  \cite{Siegel}.} 
 \cite{DF,LS,K-tensor} some time ago.
Here we wish to give an alternative formulation, as compared with (\ref{improved}),  for 
the massive improved $\cN=2$ tensor multiplet in conformal 
supergravity building on the rigid supersymmetric construction of \cite{K-tensor}. 

Let us introduce a covariantly chiral prepotential $\J$
for the tensor multiplet (see, e.g.,  \cite{Muller86} and references therein)
\bea
G^{++} (u^+)= \frac{1}{8}\Big( (\cD^{+})^2+4{S}^{++}\Big) \J
+\frac{1}{8}\Big( (\cDB^{+})^2+4\widetilde{S}^{++}\Big){\bar \J}~, \qquad
{\bar \cD}^i_\ad \J=0
\label{G++}
\eea
The prepotential is defined modulo gauge transformations of the form:
\be
\d \J ={\rm i}\, \L~, \qquad  {\bar \cD}^i_\ad \L=0~,
\qquad \Big(\cD^{\g(i}\cD_\g^{j)}+4S^{ij}\Big)\L=
\Big(\cDB_\gd^{(i}\cDB^{ j) \gd}+ 4\bar{S}^{ij}\Big)\bar{\L}~.
\ee
The super-Weyl transformation law of $\J$ should be 
\be
\d_\s \J = \s \J
\ee
in order for $G^{++}$ to transform as in eq. (\ref{super-Weyl-G}).

To describe a massive improved tensor multiplet, one can choose
the following Lagrangian:
\bea
\cL^{++} &=& -{ G}^{++} \ln \frac{{ G}^{++}}{{\rm i}\widetilde{ \U}^+ { \U}^+} \non \\
&&+\frac{1}{16}{\bm V} \,\Big\{ \m( \m+{\rm i}e)\Big( (\cD^{+})^2+4{S}^{++}\Big) \frac{\J^2}{\bm W}
~+~
\mbox{smile-conjugate}
\Big\}~,
\label{improved-massive-2}
\eea
with $\m$ and $e$ constant mass parameters. 
Here the mass terms include  both ``electric''  and ``magnetic''  contributions.
The action generated by this Lagrangian is obviously super-Weyl invariant.
The mass parameter in  (\ref{improved-massive-2}) is complex,  and
can be interpreted as a vacuum expectation value for vector multiplets\footnote{This interpretation 
is inspired by \cite{GL}.}
\bea
\m( \m+{\rm i}e)
 \int \rd^4 x \,{\rm d}^4\q \, \cE \, \J^2 \quad \longleftarrow \quad
 \int \rd^4 x \,{\rm d}^4\q \, \cE \, H(W) \J^2~,
 \eea
 with $H(W)$ a holomorphic homogeneous  function of degree zero,
with its variables $W$'s being the field strengths of  Abelian vector multiplets.

The rigid supersymmetric versions of (\ref{improved-massive-1})
and (\ref{improved-massive-2}) are dually equivalent provided the mass parameters 
are related as \cite{K-tensor}
\be
m^2 =\m^2 +e^2~.
\ee
It turns out that this duality extends to supergravity. Indeed, let us consider 
the auxiliary first-order Lagrangian:
\bea
\cL^{++}_{\rm aux}  &=& -{ U}^{++} \ln \Big( \frac{{ U}^{++}}{{\rm i}\widetilde{ \U}^+ { \U}^+} 
-1\Big)
\non \\
&&+mV \Big\{ U^{++} - \frac{1}{8}\Big( (\cD^{+})^2+4{S}^{++}\Big) \J
-\frac{1}{8}\Big( (\cDB^{+})^2+4\widetilde{S}^{++}\Big){\bar \J} \Big\} \non \\
&&+\frac{1}{16}{\bm V} \,\Big\{ \m( \m+{\rm i}e)\Big( (\cD^{+})^2+4{S}^{++}\Big) \frac{\J^2}{\bm W}
~+~
\mbox{smile-conjugate}
\Big\}~.
\eea
Here $U^{++}$ and $V$ are covariant real weight-two and weight-zero 
tropical multiplets, respectively, and $\J$ a covariantly chiral scalar.
The corresponding action, $S_{\rm aux} $, 
is invariant under the gauge transformation 
 (\ref{vm-gauge-tr}) that also acts on $\U^+$ as
$\d { \U}^+ =- m \l { \U}^+$, and hence $\U^+$ is a purely gauge degree of freedom.
Varying\footnote{The equations of motion for $\U^+$ and $\widetilde{\U}^+$
only force $U^{++}$ to be a tensor multiplet. 
The equation of motion for $V$ requires $\J$ to be the prepotential of this tensor 
multiplet.} 
$S_{\rm aux} $ with respect to $V$ leads to $U^{++}=G^{++}$, with $G^{++}$
given by eq. (\ref{G++}), and then we 
arrive at the theory with Lagrangian (\ref{improved-massive-2}).
On the other hand, varying $U^{++}$ and $\J$ can be shown to lead to 
the following model: 
\bea
\cL^{++}_{\rm dual} &=&   
-\frac{1}{16}{\bm V} \,\left\{ \Big( (\cD^{+})^2+4{S}^{++}\Big) \frac{W^2}{\bm W}
+ \Big( (\cDB^{+})^2+4\widetilde{S}^{++}\Big)\frac{{\bar W}^2}{\bar {\bm W}} \right\}
+ {\rm i} \widetilde{ \U}^+ {\rm e}^{m V}{ \U}^+~,~~~~
\label{improved-massive-3}
\eea
where $W$ is the field strength, eq. (\ref{W}), 
associated with the gauge prepotential $V$. One can further demonstrate 
that the theories (\ref{improved-massive-1}) and (\ref{improved-massive-1}), 
in complete analogy with the consideration in subsection 3.3 below. 
This completes the proof. 

In order to evaluate the variational derivatives of $S_{\rm aux} $ with respect $\J$, 
a few comments are actually in order. First of all, using the representation (\ref{W}) 
for the field strengths $\bm W$, in conjunction with integration by parts, 
one can transform the linear in $\J$ term in 
$S_{\rm aux} $ as follows:
\bea
&&-\frac{m}{16\pi}
\int \rd^4 x \,{\rm d}^4\q {\rm d}^4{\bar \q}\, E \oint (u^+ \rd u^{+})
\frac{{\bm W}{\bar {\bm W}}   }{({\bm \S}^{++})^2 }
V\Big( (\cD^{+})^2+4{S}^{++}\Big) \J \non \\
&=& \frac{m}{8\pi^2}
\int \rd^4 x \,{\rm d}^4\q {\rm d}^4{\bar \q}\,E  \oint (u^+ \rd u^{+}) \oint (\hat{u}^+ \rd {\hat u}^{+})
\frac{V(u^+) {\bm V}(\hat{u}^+)}{(\hat{u}^+u^+)^2} \, \frac{\J}{\bm W}~.\non
\eea
Relabelling here $u^+ \leftrightarrow \hat{u}^+$, then making use of the representation 
(\ref{W}) for the field strength $W$ associated with $ V$, and also integrating by parts,
the latter expression can be brought to the form:
\bea
&&-\frac{m}{16\pi}
\int \rd^4 x \,{\rm d}^4\q {\rm d}^4{\bar \q}\, E \oint (u^+ \rd u^{+})
\frac{{\bm W}{\bar {\bm W}}   }{({\bm \S}^{++})^2 }
{\bm V}\Big( (\cD^{+})^2+4{S}^{++}\Big) \frac{\J W}{\bm W}~. \non
\eea
Now, recalling the two equivalent representations (\ref{chiral-action})
for chiral actions, the $\J$-dependent terms in  $S_{\rm aux} $ become
\bea
-\frac{1}{4}  \int \rd^4 x \,{\rm d}^4\q \, \cE \, \Big\{ \m( \m+{\rm i}e)\J^2 -2m \J W\Big\}~, 
\non
\eea
and this functional is trivial to vary with respect to $\J$.

\subsection{Supergravity-matter systems with  tensor compensator}
We now turn to supergravity-matter systems. To start with,
we choose the following compensating multiplets:
(i) the vector multiplet described by its covariant real weight-zero 
{\it tropical} prepotential ${\bm V}(u^+)$, with $\bm W$  the corresponding gauge-invariant 
covariantly chiral field strength; and (ii) the  tensor multiplet ${\bm G}^{++}(u^+)$. 
As matter fields, we choose a set of covariant weight-zero arctic multiplets
$\U^I(u^+)$ and their smile-conjugates $\widetilde{\U}^{\bar I}(u^+)$ which take their 
values in a K\"ahler manifold, with $K(\upsilon^I,{\bar \u}^{\bar J} )$
the corresponding K\"ahler potential. The supergravity-matter  Lagrangian is
\bea
\cL^{++}_{\rm SM-tensor} = \hf {\bm V}\,{\bm \S}^{++}+
{\bm G}^{++} \Big(
\ln \frac{{\bm G}^{++}}{{\rm i}\widetilde{ \U}^+{\rm e}^{m\bm V}{ \U}^+} -K(\U, \widetilde{\U}) \Big)~,
\label{SM1}
\eea
with $m$ a cosmological constant.
It should be pointed out that the vector
and the tensor multiplet  kinetic terms appear here with  wrong signs,
as compared with  (\ref{vm-lagrangian}) and (\ref{improved}).
Since $K(\u^I,{\bar \u}^{\bar J} )$ is essentially arbitrary,
and the covariant polar multiplets were discovered in \cite{KLRT-M},  
the theory introduced describes more general matter couplings
than previously constructed
within the third minimal formulation for $\cN=2$ Poincar\'e supergravity \cite{deWPV}.

The action generated by (\ref{SM1}) is invariant under the gauge
transformations of the compensating vector multiplet,
eq. (\ref{vm-gauge-tr}). It is also super-Weyl invariant, since 
the covariant weight-zero projective multiplets are invariant under such 
transformations \cite{KLRT-M}, 
\be
\d_\s \U^I =0~.
\ee
In addition, the action possesses the K\"ahler invariance 
\be
K(\U, \widetilde{\U}) \to K(\U, \widetilde{\U}) +\L(\U) +{\bar \L}(\widetilde{\U})~,
\label{Kahler}
\ee
with $\L$ an arbitrary holomorphic function.

It is interesting to note that the equation of motion for $\bm V$ is 
\be
{\bm \S}^{++}+ m {\bm G}^{++}=0~.
\ee
Then, the dynamics of the matter sector is generated by
the Lagrangian
\be
\cL^{++} \propto  {\bm \S}^{++} K(\U, \widetilde{\U}) ~.
\ee
A similar model has been introduced in \cite{KT-Msugra3}
in the case of 5D $\cN=1$ supergravity.

\subsection{Supergravity-matter systems with  polar compensator}

Let us derive a dual formulation for the theory (\ref{SM1}).
Instead of (\ref{SM1}),
we consider the following first-order Lagrangian:
\bea
\cL^{++}_{\rm first-order} = \hf {\bm V}\,{\bm \S}^{++}+
U^{++} \Big(
\ln \frac{U^{++}}{{\rm i}\widetilde{{\bm \U}}^+{\rm e}^{m\bm V}{\bm  \U}^+}
 -1-K(\U, \widetilde{\U}) \Big)~.
\label{first-order}
\eea
Here $U^{++}$ is a covaraint real weight-two {\it tropical} multiplet,
and ${\bm \U}^+$ a covariant weight-one arctic multiplet.
Unlike the purely gauge  superfield ${ \U}^+$ in the original model (\ref{SM1}), 
the ${\bm \U}^+$ is now a non-trivial dynamical variable.
The first-order model introduced respects all the symmetries of the 
original theory (\ref{SM1}), albeit in a modified form. The gauge invariance 
 (\ref{vm-gauge-tr}) turns into 
\be
\d {\bm V} =\l  + \widetilde{\l}~, \qquad \d {\bm \U}^+ =- m \l {\bm \U}^+~.
\label{vm-gauge-tr2}
\ee
The K\"ahler transformation (\ref{Kahler}) becomes
\be
K(\U, \widetilde{\U}) \to K(\U, \widetilde{\U}) +\L(\U) +{\bar \L}(\widetilde{\U})~,
\qquad {\bm \U}^+ \to {\rm e}^{-\L(\U) }{\bm \U}^+.
\label{Kahler2}
\ee
Finally, if the super-Weyl transformation of $U^{++} $ is chosen to be 
\be
\d_\s U^{++} = (\s +{\bar \s})U^{++}~,
\ee
then the action $S_{\rm first-order}$ associated with (\ref{first-order}) is 
super-Weyl invariant. 

Varying $S_{\rm first-order}$ with respect to ${\bm \U}^+$ and 
$\widetilde{\bm \U}^+$ constrains $U^{++}$ to be  an $O(2)$-multiplet,  
$U^{++} =G^{++}$, and then $S_{\rm first-order}$ reduces to the action 
generated by (\ref{SM1}). Therefore, the dynamical systems 
(\ref{SM1}) and (\ref{first-order}) are equivalent.
On the other hand, varying $S_{\rm first-order}$ with respect to $U^{++}$
leads to the following theory:
\bea
\cL^{++}_{\rm SM-polar} =\hf  {\bm V}\,{\bm \S}^{++} 
- {\rm i} \,\widetilde{\bm \U}^+ {\rm e}^{m {\bm V} - K(\U, \widetilde{\U})}{\bm \U}^+~.
\label{SM2}
\eea
This is exactly the supergravity-matter system introduced in \cite{KLRT-M}.
Its hypermultiplet sector is a curved-space version of the general 
4D $\cN=2$ superconformal sigma-model for polar multiplets 
proposed in \cite{K-hyper2} (building on the 5D $\cN=1$ construction 
of \cite{K-hyper1}).

It is instructive  to compare (\ref{SM1}) and (\ref{SM2})
to the well-known descriptions of matter couplings in the
new minimal and the old minimal formulations 
for $\cN=1$ supergravity (see, e.g., \cite{BK} for a review).
The action for the matter-coupled new minimal supergravity is
as follows:
\be
\label{sigma}
S=  \int \rd^4 x \,{\rm d}^2\q {\rm d}^2{\bar \q}\, E\, \Big\{
3G\, {\rm ln} \frac{G}{\j \bar \j} + 
{G}\,K\!(\f, \bar \f )\Big\}~.
\ee
Here the compensator $G$ is covariantly real linear,
and $\j$ is a covariantly chiral scalar which is a pure gauge degree of freedom.
The supersymmetric matter is described by covariantly chiral scalars 
$\f^I$ which are inert under the super-Weyl transformations.
The action is super-Weyl invariant, and also possesses the K\"ahler invariance
\be
\label{kahler}
K\!(\f, \bar \f) \to K\!(\f, \bar \f) + \L(\f) + \bar \L(\bar \f)~,
\ee
with $\L(\f)$ a holomorphic function. 
The action for the matter-coupled old minimal supergravity is
\be
\label{kahlermodel}
S = -3 \int \rd^4 x \,{\rm d}^2\q {\rm d}^2{\bar \q}\, E\,
{\bar \J}\, \J \, 
{\rm exp}\!\left(-\frac 13 K\!(\f,{\bar \f})\right)~,
\ee
where the compensator  $\J$ is  covariantly chiral. 
The two theories (\ref{sigma}) and (\ref{kahlermodel}) are dually 
equivalent. Clearly, the $\cN=2$ supergravity theories 
(\ref{SM1}) and (\ref{SM2}) are generalizations of 
(\ref{sigma}) and (\ref{kahlermodel}), respectively.

In principle, the supergravity-matter systems  (\ref{SM1}) and (\ref{SM2})
can be used to generate arbitrary quaternion-K\"ahler geometries
allowed in supergravity. In this respect, the results of \cite{deWRV}
should also be relevant.

Ten years ago, it was advocated in \cite{K-double} that the harmonic \cite{GIKOS,GIOS}
and projective \cite{KLR,LR} superspace approaches provide complementary 
descriptions of rigid $\cN=2$ supersymmetric theories. This  also  appears to hold at 
the level of $\cN=2$ supergravity. The strong features of the projective-superspace  
formulation  proposed in \cite{KLRT-M} 
are: (i) its geometric character; (ii) reasonably short off-shell hypermultiplets.
The strongest point of the harmonic-superspace approach to $\cN=2$
supergravity \cite{GIOS2,GIOS} is a remarkably simple structure of the 
supergravity prepotentials.
In particular, the latter  approach provides 
a natural origin for the real scalar prepotenial that generates     
the $\cN=2$ supercurrent \cite{Siegel-curved,KT}. It would be important to understand
how such a prepotential originates within the projective-superspace scheme.
\\

\noindent
{\bf Acknowledgements:}\\
I am very grateful to  Gabriele Tartaglino-Mazzucchelli for collaboration 
at early stage of this project and for comments on the manuscript.
I also thank Ian McArthur for reading the manuscript.
The hospitality of the 2008 Simons Workshop in Mathematics and Physics,
 where this project was completed, is gratefully acknowledged.
The research presented in this work is supported 
by the Australian Research Council.

\appendix

\section{Equivalent forms for the vector multiplet action} 
\setcounter{equation}{0}
Here we list   four equivalent forms for the free vector multiplet action:
\begin{subequations}
\bea
S_{\rm VM} & = & \hf \int \rd^4 x \,{\rm d}^4\q \, \cE \, W^2 
=\frac{1}{4} \int \rd^4 x \,{\rm d}^4\q \, \cE \, W^2 ~+~{\rm c.c.} 
\label{form1} \\
&=& \frac{1}{2\pi} \oint (u^+ \rd u^{+})
\int \rd^4 x \,{\rm d}^4\q {\rm d}^4{\bar \q}\, E\,
\frac{{\bm W}{\bar {\bm W}}  }{({\bm \S}^{++})^2 } \,\cL^{++}_1 
\label{form2} \\
&=& \frac{1}{2\pi} \oint (u^+ \rd u^{+})
\int \rd^4 x \,{\rm d}^4\q {\rm d}^4{\bar \q}\, E\,
\frac{{\bm W}{\bar {\bm W}}  }{({\bm \S}^{++})^2 } \,\cL^{++}_2
\label{form3}  \\
&=&\hf \frac{1}{(2\pi)^2} \oint (u^+_1 \rd u^{+}_1)  \oint (u^+_2 \rd u^{+}_2)
\int \rd^4 x \,{\rm d}^4\q{\rm d}^4{\bar \q}\,E\,
\frac{{ V}(u^+_1){ V}(u^+_2)}{(u^+_1u^+_2)^2}~,
\label{form4} 
\eea
\end{subequations}
where
\begin{subequations}
\bea
\cL^{++}_1 &=& -\hf { V}\,{ \S}^{++}~, \\
\cL^{++}_2 &=&
 -\frac{1}{16} {\bm V} \,\left\{ \Big( (\cD^{+})^2+4{S}^{++}\Big) \frac{W^2}{\bm W}
+\Big( (\cDB^{+})^2+4\widetilde{S}^{++}\Big)
\frac{{\bar W}^2 }{\bar {\bm W}} \right\}~.
\eea
\end{subequations}
Let us derive, for instance, (\ref{form4}) from (\ref{form3}). 
It is sufficient to consider the $W^2$-dependent part of (\ref{form3}). 
Integrating by parts using the representation (\ref{W}) 
for the field strengths $ W$, and  integrating by parts once more,
one can show that
\bea
&-&\frac{1}{32\pi}
\int \rd^4 x \,{\rm d}^4\q {\rm d}^4{\bar \q}\, E \oint (u^+ \rd u^{+})
\frac{{\bm W}{\bar {\bm W}}   }{({\bm \S}^{++})^2 }
{\bm V} \Big( (\cD^{+})^2+4{S}^{++}\Big) \frac{W^2}{\bm W} \non \\
&=& \frac{1}{16\pi^2}
\int \rd^4 x \,{\rm d}^4\q {\rm d}^4{\bar \q}\,E  \oint (u^+ \rd u^{+}) \oint (\hat{u}^+ \rd {\hat u}^{+})
 \frac{W}{\bm W}\,
\frac{{\bm V}(u^+) { V}(\hat{u}^+)}{(\hat{u}^+u^+)^2} ~.
\non
\eea
As a next step, one can re-label $u^+ \leftrightarrow \hat{u}^+$ in the expression obtained, 
insert the unity resolved as ${\bm \S}^{++} / {\bm \S}^{++}$,  and then integrate by parts, 
thus ending up with 
\bea
-\frac{1}{8\pi}
\int \rd^4 x \,{\rm d}^4\q {\rm d}^4{\bar \q}\,E  \oint (u^+ \rd u^{+}) 
\frac{ {\bar {\bm W} } }{ {\bm \S}^{++} } V \,W~.
\non
\eea
It remains to make use, once more,  of the representation (\ref{W}) 
for the field strengths $ W$, and then integrate by part, in order to 
arrive at (\ref{form4}).

\end{document}